\documentclass[aps,reprint,graphicx]{revtex4-1}

\usepackage{graphicx}
\usepackage{epstopdf}
\usepackage{dcolumn}
\usepackage{bm}
\usepackage{amsmath}
\usepackage{lipsum}

\newcommand{\fref}[1]{Figure \ref{#1}}
\newcommand{\eref}[1]{(\ref{#1})}

\draft 

\begin{document}

\title{Mode localization and sensitivity  in weakly coupled resonators}

\author{M. Manav}
\affiliation{Mechanical Engineering, University of British Columbia, Vancouver, Canada}

\author{A. S. Phani}
\email{srikanth@mech.ubc.ca}
\affiliation{Mechanical Engineering, University of British Columbia, Vancouver, Canada}

\author{E. Cretu}
\email{edmondc@ece.ubc.ca}
\affiliation{Electrical and Computer Engineering, University of British Columbia, Vancouver, Canada}

\date{\today}

\begin{abstract}
Localization of normal modes  is used in recent  microelectromechanical systems (MEMS)  technologies with orders of magnitude improvements in sensitivity.  A  pair of eigenvalues  veer, or avoid crossing each other, as a \emph{single} parameter of a vibrating system is varied. While it is well-known that the sensitivity ($s$) of  modal  amplitude  ratio varies with strength of coupling ($\kappa$) as $s \propto\kappa^{-1} $ in the case of two identical coupled oscillators, recently, we showed  that  asymmetry $\alpha$  will also influence sensitivity according to $s \propto (\alpha \kappa)^{-1}$.  Here, we show  that further enhancements in sensitivity is possible in higher degrees of freedom ($n$) systems using energy analysis. In the case of $n-2$ uniformly coupled oscillators embedded between two  oscillators, we show that  $s \propto  \alpha^{-1}\kappa^{1-n}$, if the \emph{blocked} resonance spectra of the embedded oscillators and the end oscillators are well-separated. We also show that asymmetric coupled oscillators also enhance linear range in addition to sensitivity when compared to their symmetric counterparts. We do not use a perturbation approach in our energy analysis;  hence the sensitivity and linear range expressions derived have a wider range of accuracy.

 \end{abstract}

\maketitle 

\section{Introduction}
\begin{figure}[!h]
	\center
	\includegraphics[width=8cm]{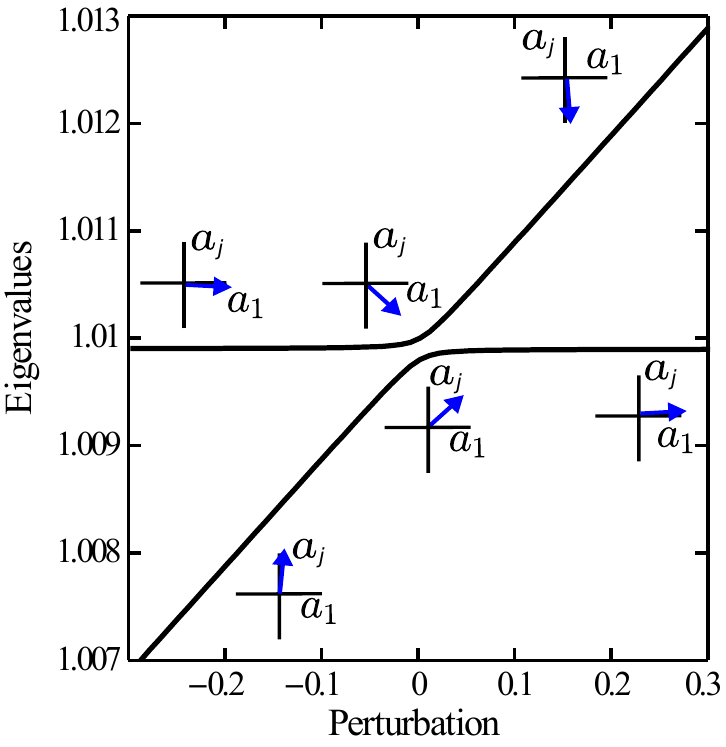}
	\caption{\label{fig:SchematicVeering} Schematic showing veering of eigenvalues (solid lines) as a system parameter is varied (only veering branches are shown). Mode localization to $j_{th}$ resonator is also shown, where $a_1$ and $a_j$ are modal amplitudes of first and $j_{th}$ resonators respectively.}
\end{figure}

Natural frequencies and mode-shapes of a linear vibrating system can  exhibit  startling sensitvitiy when a parameter is varied.  Stated mathematically, eigenvalues (square of undamped natural frequencies) and eigenvectors (normal modeshapes) are sensitive to a parameter change in the underlying matrix  differential operator. How these quantities change in the vicinity of a degenerate point has attracted the attention of many physicists~\cite{teller37,arnol89} and engineers~\cite{warburton54,nair73,hodges1982,hodges1983,perkins1986,pierre1988,mace12}.   In eigenvalue veering, two eigenvalue curves come close as a parameter of a linear vibrating system (for example mass, or stiffness) is varied.  Instead of crossing, as one anticipates, they veer away from each other as shown in~\fref{fig:SchematicVeering}. Simultaneously, the associated eigenvectors with each curve rotate, culminating in localization of vibrational energy to specific resonators. Given the fundamental nature of eigenvalue problems, it is not surprising that this phenomenon of veering or avoided crossing has been studied, often independently, in various branches of  physics, structural dynamics and musical acoustics~\cite{woodhouse2004}. Here, our concern is with  the phenomenon of localization of normal modes of a discrete coupled linear vibrating system in the context of sensing.

Rapid developments in miniaturization technologies for  sensors, coupled with the need for higher sensitivity and the search for alternatives to conventional  resonant frequency shift-based sensing paradigms used in an atomic force microscope~\cite{albrecht91} for example, has revived interest in mode localization as a means to achieve ultra high sensitivity~\cite{spletzer06,thiruvenkatanathan09,zhao2016mems}.  Orders of magnitude improvements in sensitivity have been achieved and novel MEMs sensors for  sensing mass~\cite{spletzer06,spletzer08,thiruvenkatanathanMass}, displacement~\cite{thiruvenkatanathanDisp}, acceleration~\cite{zhangAcceleration,yangAcceleration}, electrometer~\cite{thiruvenkatanathanElectrometer,zhangElectrometer} have emerged.  The eigenvector sensitivity is exploited in these sensing technologies. Within a narrow range of perturbation, called veering zone, the eigenvectors rotate swiftly~\cite{du2011,manav2014,manav2017} and the rate of rotation is inversely proportional to coupling stiffness ($\kappa$). Consequently, sensitivity ($s$) of  eigenvectors to a mass or stiffness perturbation is very high for weak coupling ( $s \propto \kappa^{-1}$, and $\kappa <<1$ for weak coupling). Typically, electrostatic coupling of resonators~\cite{thiruvenkatanathan09} yields a weak, tuneable coupling stiffness leading to high sensitivity. Several transducers that use electrostatic coupling springs have been reported based on mass perturbation~\cite{thiruvenkatanathanMass} or stiffness perturbation~\cite{thiruvenkatanathan09,thiruvenkatanathanDisp,manav2014}, operating in vacuum~\cite{thiruvenkatanathan09,thiruvenkatanathanDisp} or in ambient conditions~\cite{manav2014,manav2017}. Closed-loop accelerometers have also been reported~\cite{kang2018closed,yang2018closed}. Contrary to common belief, symmetry of resonators is not a prerequisite for veering and mode localization to occur~\cite{stephen2009,zhao2016mems,manav2017}.   In fact, when two asymmetric  resonators are coupled, the sensitivity of modal amplitudes undergoing veering varies as  $s \propto (\alpha\kappa)^{-1}$, where $\alpha$ is degree of asymmetry. Thus, indeed asymmetry can help improve sensitivity for $\alpha<1$.  Here, we show using energy analysis that further enhancements in sensitivity are possible by increasing the order of the vibrating system and we also quantify the trade-off  between sensitivity and bandwidth.

We begin by describing a theoretical framework based on energy and exploit synchronous motion properties to deduce recursive \emph{modal} equations to explain mode localization in section~\ref{theory}, followed by a derivation of approximate sensitivity expression for an $n$-DOF coupled resonator system in section~\ref{N-DOF_analysis}. We then specialize to the case of a  three coupled resonator system in section~\ref{3-DOF_analysis}, ending with conclusions in section~\ref{conclusion}.

\section{Analysis of veering}
\label{theory}
\begin{figure}
	\center
	\includegraphics[width=8cm]{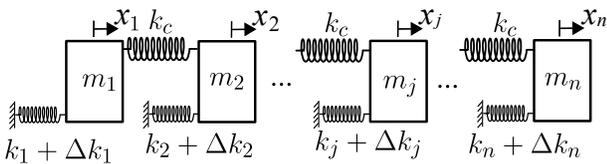}
	\caption{\label{fig:SchematicDOFn} An $n$-DOF coupled spring-mass system with nonuniform stiffness perturbation.}\label{fig2}
\end{figure}
In this section, we establish relation between modal amplitudes of an $n$-DOF coupled resonators system shown in \fref{fig:SchematicDOFn}. $m_j$, $k_j$ and $\Delta k_j$ are mass, stiffness and perturbation in stiffness respectively of $j_{th}$ resonator. Resonators are coupled using springs of stiffness $k_c$. Traditionally one uses \emph{matrix} based linear algebraic principles to analyze mode localization, such as matrix based perturbation method~\cite{courant1965} or eigenderivative method~\cite{fox68,pierre1988,du2011}. Here, we use a general energy based approach to develop recursive relations for \emph{modal} amplitudes. The advantage of this approach is that the perturbations need not be small. 

We start by writing governing equations for the system shown in~\fref{fig2}. To find the amplitudes of the normal modes from the governing equations, we employ the concept of synchronous motion. In synchronous motion, all the masses pass through their respective equilibrium (as well as minima and maxima) positions at the same time~\cite{rosenberg60,rosenberg62}. Note that for an undamped or proportionally damped system, synchronicity is a property of normal modes. Although the analysis below assumes a conservative system (no damping), the modal relations obtained in the end are valid for a proportionally damped system as well~\cite{phani2003,phani2007}.  

The governing equations of motion for the conservative system are given by:
\begin{equation}
	m_j\ddot{x}_j =-\frac{\partial V}{\partial x_j},
	\label{eq:eq1}
\end{equation}
where $x_j$ is displacement of mass $m_j$ and $V$ is the potential energy of the system. Potential energy can come from strain, electrostatic or magnetic interactions etc. or a combination of those. Here we assume them to be stored in springs of constant stiffnesses (linearly vibrating system). Then the potential energy expression is obtained as:
\begin{equation}
	V=\frac{1}{2}\sum\limits_{i=1}^n{(k_i+\Delta k_i)x_i^2}+\frac{1}{2}k_c\sum\limits_{i=1}^{n-1}{(x_{i+1}-x_i)^2}.
	\label{eq:eq2}
\end{equation}
We refer to~\fref{fig2} for the definition of parameters used in the above equation. Now, for a generalized synchronous motion, displacement of one mass is sufficient to characterize the motion of the system as all displacements are algebraically related~\cite{rosenberg60,rosenberg62}. For a linear system, the synchronous motion is of the following form:
\begin{equation}
	x_j(t)=a_j x_1(t),
	\label{eq:eq3}
\end{equation}
where we take the first mass as the reference mass (assuming $x_1\neq 0$, $a_1=1$). $a_j$ are constant modal amplitude parameters. Note that we have selected the first mass as reference with unit amplitude of the mode, $a_1=1$. Thus for all modes $p=1\hdots n$, $a_{1p}=1$. The normal modal vector of the system can be written as $[1~a_2~ ...~ a_n]^T$. Substituting \eref{eq:eq2} and \eref{eq:eq3} in \eref{eq:eq1}, the following recursive relations among modal amplitude parameters is obtained~\cite{manav2014}:
\begin{eqnarray}
	a_{j-1}(1-\delta_{j1})\nonumber \\
	\resizebox{0.43\textwidth}{!}{$-\left (\frac{k_j-\alpha_j k_1}{k_c}+\frac{\Delta k_j-\alpha_j\Delta k_1}{k_c}+(2-\alpha_j-\delta_{j1}-\delta_{jn})+\alpha_j a_2 \right )a_j$}\nonumber \\
	+a_{j+1}(1-\delta_{jn})=0 \label{eq:eq3b}.
\end{eqnarray}
where $\alpha_j=m_j/m_1$ is the \emph{asymmetry} parameter, $j=1...n$. $\delta_{ij}$ is a Kronecker delta function.  The above set of \emph{recursive} relations simplify further if the perturbation $\Delta k_j$ is localized. For perturbation only in the $l_{th}$ resonator ($\Delta k_j=\delta_{jl}\Delta k$) with $l\ne1$ (no perturbation in the reference resonator), the recursive relations reduce to the following form:
\begin{eqnarray}
	a_{j-1}(1-\delta_{j1})\nonumber \\
	\resizebox{0.43\textwidth}{!}{$-\left (\frac{k_j-\alpha_j k_1}{k_c}+\frac{\Delta k}{k_c}\delta _{jl}+(2-\alpha_j-\delta_{j1}-\delta_{jn})+\alpha_j a_2 \right )a_j$}\nonumber \\
	+a_{j+1}(1-\delta_{jn})=0 \label{eq:eq4}.
\end{eqnarray}

The recursive relations in~\eref{eq:eq4} can be transformed into an $n_{th}$ order polynomial equation in $a_2$ alone, in principle. The $n$ roots of this polynomial equation, $a_{2p},~p=1\ldots n$, then are the modal amplitudes of the second mass in each of the $n$ modes of the system. Substituting the values of $a_2=a_{2p}$ in the recursive relation \eref{eq:eq4} gives the modal amplitudes of other (third, fourth etc.) resonators in the $p_{th}$ mode. Further, the $p_{th}$ natural frequency is obtained by inserting simple harmonic motion of the resonators, $x_1(t)=a_1e^{i\omega_p t}=e^{i\omega_p t}$ and $x_2(t)=a_2e^{i\omega_p t}$ in the governing equation for the first mass (the first equation in \eref{eq:eq1} corresponding to $j=1$):
\begin{eqnarray}
	\omega_p=\sqrt{\frac{k_1+\Delta k_1+k_c(1-a_{2p})}{m_1}}, \label{eq:eq5}
\end{eqnarray}
where $a_{2p}$ is the modal amplitude of the second mass in $p_{th}$ mode of vibration. Note that $p_{th}$ eigenvalue for this system equals square of $p_{th}$ natural frequency. In order to characterize mode localization in $p_{th}$ mode due to stiffness perturbation, we define the following eigenmode \emph{sensitivity} norm ($S_p$):
\begin{equation}
	S_p=\frac{da_{np}}{d\delta},  \quad \delta=\Delta k/k_1. \label{eq:eq6}
\end{equation}
where $a_{np}$ is modal amplitude \emph{ratio} since $a_{1p}=1$, and $\delta$ is nondimensional stiffness perturbation. We need to interpret $a_{np}$ as the ratio of displacement amplitude of the $n_{th}$ resonator to the first resonator.  Note that nondimensional stiffness perturbation $\delta$ is different from Kronecker delta $\delta_{ij}$. For latter use, we define nondimensional coupling stiffness as $\kappa=k_c/k_1$.

\section{Sensitivity of $n$-coupled resonators}
\label{N-DOF_analysis}
Here we derive an expression for the eigenmode sensitivity, as defined in \eref{eq:eq6}, to stiffness perturbation in the $n_{th}$ resonator in a system of $n$ weakly coupled resonators, where the first and the last resonators have the same natural frequency ($\omega_0^2=\frac{k_1}{m_1}=\frac{k_n}{m_n}$), but the middle resonators have very high natural frequencies compared to the two end resonators ($\omega_j>>\omega_0, j=2\hdots n-1$). Having resonators of the same natural frequency at the two ends is guided by the fact that the stiffer resonators in the middle block transmission of energy across the coupled resonators system due to impedance mismatch. This leads to very high ratio of modal amplitudes of the end resonators in a veering mode, and hence enhanced energy localization in a vibration transmission when one end of the chain is harmonically forced.

\begin{figure}[!h]
	\center
	\includegraphics[width=8cm]{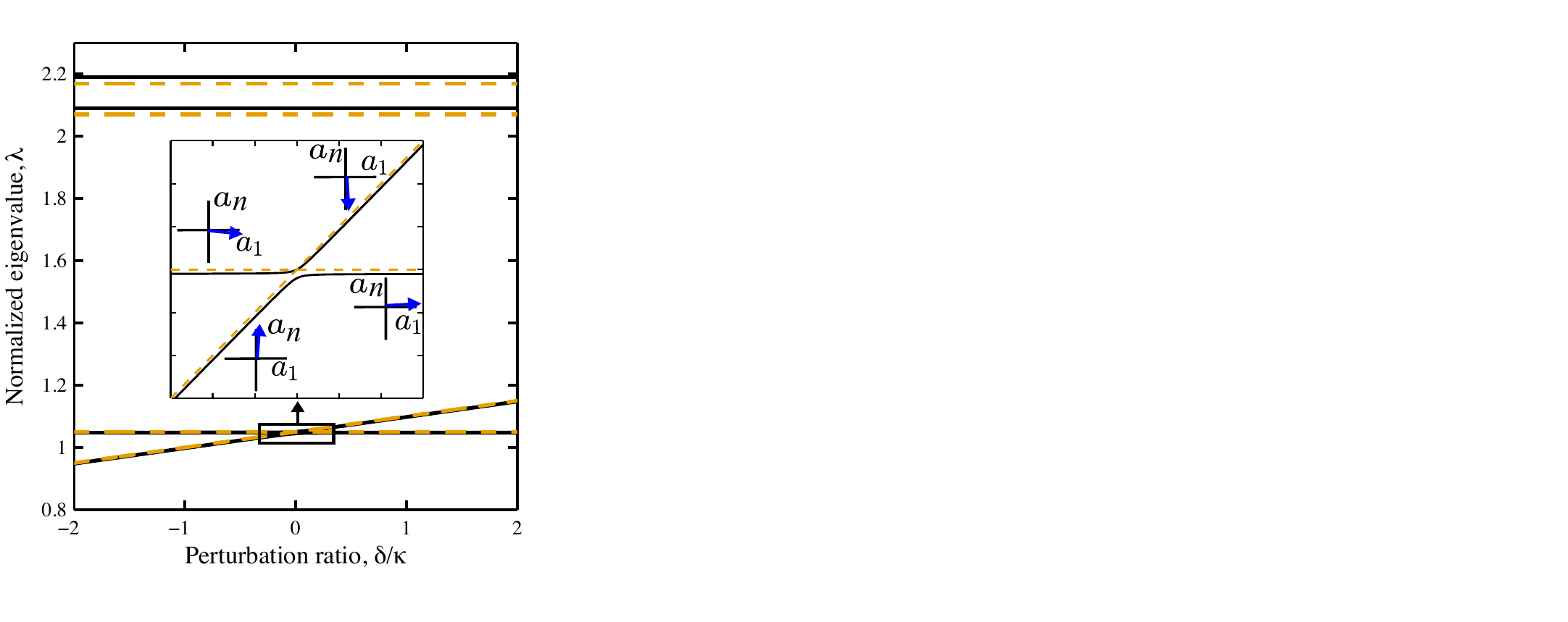}
	\caption{\label{fig:SchematicVeeringDOFn} An schematic showing square of blocked uncoupled natural frequencies (dashed line) and square of natural frequencies of the coupled system (solid line) as stiffness perturbation in the last resonator is varied in an $n$-DOF system (only first four branches are shown here) with middle resonators of very high natural frequency compared to the two end resonators.}
\end{figure}

Consider the variation of normalized eigenvalues, $\lambda_p=\omega_p^2/\omega_0^2,~\omega_0^2=k_1/m_1$, of the system in~\fref{fig:SchematicVeeringDOFn} as stiffness perturbation in the last resonator, denoted by perturbation ratio ($\Delta k/k_c=\delta/\kappa$), is changed. The plot also shows the variation of the square of the blocked uncoupled natural frequencies (motion of all the resonators are blocked except the one in consideration)~\cite{mace12} of each of the resonators as dashed lines. The frequency curves which cross in blocked uncoupled case veer in the coupled system, whereas the higher frequencies undergo negligible change, see~\fref{fig:SchematicVeeringDOFn}. Veering frequencies asymptotically approach blocked uncoupled frequencies~\cite{mace12} and hence can be approximated by them \emph{outside} the veering zone. For positive perturbations, $\delta/\kappa>0$, and we have
\begin{eqnarray}
	\lambda_2\approx\frac{k_n+k_c+\Delta k}{m_n}\approx\omega_0^2\left(1+\frac{\kappa}{\alpha_n}+\frac{\delta}{\alpha_n}\right),\nonumber \\
	\omega_0^2=\frac{k_1}{m_1}=\frac{k_n}{m_n},~\kappa=\frac{k_c}{k_1},~\delta=\frac{\Delta k}{k_1},~\alpha_n=\frac{m_n}{m_1},
	\label{eq7}
\end{eqnarray}
Though having the same natural frequency of the first and the last resonators is sufficient, the two resonators have been assumed to be identical, i.e. $k_1=k_n,~m_1=m_n$ to ease subsequent derivation, without loosing generality of the conclusions. Thus the asymmetry parameter $\alpha_n=1$. The above expression can also be obtained through natural frequency approximation using Rayleigh quotient method~\cite{phani2008} ($\lambda_2=\frac{\phi^TK\phi}{\phi^TM\phi}$, where $\phi$ is modeshape, and $K$ and $M$ are stiffness and mass matrices) by assuming that the mode is localized to the last resonator ($\phi=[0~0~...~0~1]^T$). Using \eref{eq:eq5} and \eref{eq7}, $a_{22}$, the modal amplitude ratio for the second mode at resonator 2, can be obtained as:
\begin{equation}
	a_{22}=1-\left(\frac{\lambda_2}{\omega_0^2}-1\right)/\kappa\approx -\frac{\delta}{\kappa}.
	\label{eq8}
\end{equation}
The modal amplitude relations \eref{eq:eq4} for $n$-coupled resonators, for the second mode ($j=2$), with stiffness perturbation only in the end resonator can be expanded as follows:
\begin{eqnarray}
	\resizebox{0.41\textwidth}{!}{$1-\left (\frac{k_2/k_1-\alpha_2}{\kappa}+(2-\alpha_2)+\alpha_2 a_2 \right )a_2+a_{3}=0,$} \nonumber \\
	\resizebox{0.41\textwidth}{!}{$a_{s-1}-\left (\frac{k_s/k_1-\alpha_s}{\kappa}+(2-\alpha_s)+\alpha_s a_2 \right )a_s+a_{s+1}=0,$}\nonumber \\
	s=3~...~N-1, \nonumber \\
	\resizebox{0.41\textwidth}{!}{$a_{n-1}-\left (\frac{k_n/k_1-\alpha_n}{\kappa}+\frac{\delta}{\kappa}+(2-\alpha_n)+\alpha_n a_2 \right )a_n=0,$}
	\label{eq9}
\end{eqnarray}
where we have used the relations $\kappa=k_c/k_1$ and $\delta=\Delta k/k_1$.

As the veering natural frequencies are much smaller than the other natural frequencies, outside the veering zone, $\lambda_1<\lambda_2<<\lambda_p,~p=3~...~n$ for $\delta>0$. This yields:
\begin{equation}
	\delta<<\frac{k_j/k_1-\alpha_s}{\alpha_s},~s=2\ldots n-1. \label{eq9b}
\end{equation}

Using the above approximation in \eref{eq9}, we get:
\begin{equation}
	\frac{k_s/k_1-\alpha_s}{\kappa}+(2-\alpha_s)+\alpha_sa_2\approx\frac{k_s/k_1-\alpha_s}{\kappa},~s=3~...~N-1 \label{eq9c}
\end{equation}
for weak coupling. Applying this approximation in the recursive relation \eref{eq9}, we obtain the modal amplitudes of the resonators in the second mode:
\begin{eqnarray}
	a_{22}\approx-\frac{\delta}{\kappa}, \nonumber \\
	a_{32}\approx-\frac{\delta}{\kappa^2}(k_2/k_1-\alpha_2), \nonumber \\
	.\nonumber \\
	.\nonumber \\
	a_{n2}\approx-\frac{\delta}{\kappa^{n-1}}\prod_{i=2}^{n-1}(k_i/k_1-\alpha_i). \label{eq10}
\end{eqnarray}
Sensitivity of the second mode (one of the modes undergoing veering) is given by:
\begin{eqnarray}
	S_2=\frac{da_{n2}}{d\delta}\approx-\frac{1}{\kappa^{n-1}}\prod_{i=2}^{n-1}(k_i/k_1-\alpha_i)\nonumber \\
	\approx-\frac{1}{\kappa^{n-1}}\left(\prod_{i=2}^{n-1}\alpha_i\right)\prod_{p=3}^{n}(\lambda_p-1),
	\label{eq11}
\end{eqnarray}
upon approximating $p_{th}$ natural frequency to be equal to the blocked uncoupled natural frequency of $i_{th}$ resonator ($\omega_p^2\approx (k_i+2k_c)/m_i\approx k_i/m_i$). The above suggests that the sensitivity increases as $1/\kappa^{n-1}$ and is also affected by the distance between the veering eigen branches and the higher eigen branches shown in \fref{fig:SchematicVeeringDOFn}. For, $k_n/k_1=m_n/m_1=\alpha_n$, it can be shown that
\begin{eqnarray}
	S_2=\frac{da_{n2}}{d\delta}\approx-\frac{1}{\alpha_n\kappa^{n-1}}\prod_{i=2}^{n-1}(k_i/k_1-\alpha_i)\nonumber \\
	\approx-\frac{1}{\alpha_n\kappa^{n-1}}\left(\prod_{i=2}^{n-1}\alpha_i\right)\prod_{p=3}^{n}(\lambda_p-1). \label{eq12}
\end{eqnarray}
The expression above suggests that the sensitivity varies inversely with $\alpha_n$. So, a smaller last resonator (also the resonator which is perturbed) having the same natural frequency as the first resonator leads to a further enhancement in sensitivity. Also, asymmetry parameters can be adjusted to improve sensitivity.

\section{Three-coupled resonators}
\label{3-DOF_analysis}
Now we deduce the amplitude ratio relation and eigensensitivity for a three-DOF coupled resonators system using \eref{eq:eq4} and compare with the approximate expression in \eref{eq12}. In the previous section, we started by finding the approximate value of one of the veering natural frequencies. Here, we start by finding the approximate value of the natural frequency branch not participating in veering. It will be seen later in this section that this allows approximation of the lowest stiffness perturbation at which variation of the modal amplitude ratio with stiffness perturbation becomes linear.

For a weakly coupled resonators system having end resonators of the same natural frequency, the middle resonator of much higher natural frequency, and the stiffness perturbation ($\Delta k$) applied to the end resonator, the modal amplitude relations \eref{eq:eq4} are as follows:
\begin{eqnarray}
	1-\left (\frac{k_2/k_1-\alpha_2}{\kappa}+2-\alpha_2+\alpha_2a_2 \right )a_2+a_3=0, \label{eq:eq14a}\\
	a_2-\left (\frac{\delta}{\kappa}+1-\alpha_3+\alpha_3a_2 \right )a_3=0. \label{eq:eq14b}
\end{eqnarray}
Eliminating $a_3$ from the recursive relation above, gives a cubic equation in $a_2$:
\begin{eqnarray}
	\resizebox{0.41\textwidth}{!}{$\alpha_2\alpha_3a_2^3+\left (\alpha_3\left(\frac{k_2/k_1-\alpha_2}{\kappa}+2-\alpha_2\right)+\alpha_2\left(1-\alpha_3\right) \right )a_2^2-$} \nonumber \\
	\resizebox{0.41\textwidth}{!}{$\left (1+\alpha_3-\left(\frac{k_2/k_1-\alpha_2}{\kappa}+2-\alpha_2\right)\left(1-\alpha_3\right)\right )a_2+\alpha_3-1+$} \nonumber \\
	\frac{\delta}{\kappa}\left(\alpha_2a_2^2+\left(\frac{k_2/k_1-\alpha_2}{\kappa}+2-\alpha_2\right)a_2-1\right)=0. \label{eq:eq14c}
\end{eqnarray}
We solve this cubic equation approximately to find $a_{21}$, $a_{22}$ and $a_{23}$, the values of modal amplitude of resonator two in the three modes of vibration of the system, using blocked uncoupled natural frequency of the middle resonator. Note that the blocked uncoupled natural frequency of the middle resonator is far away from the veering natural frequencies and is approximately equal to the third natural frequency of the system (see \fref{fig:SchematicVeeringDOFn}, where for a 3-DOF system, the topmost eigen branch will be absent). Using this approximation, we get:
\begin{equation}
	\omega_3\approx\sqrt{\frac{k_2+2k_c}{m_2}}. \label{eq:eq14d}
\end{equation}
Using \eref{eq:eq5}, amplitude of the second mass in the third mode of vibration ($a_{23}$) is approximated.
\begin{equation}
	a_{23}\approx 1-\frac{k_2-\alpha_2k_1+2k_c}{\alpha_2k_c}\approx -\frac{k_2/k_1-\alpha_2+2\kappa-\alpha_2\kappa}{\alpha_2\kappa}. \label{eq:eq14e}
\end{equation}
Note that $a_{23}$ is large as $\kappa<<1$. Now, as $a_{23}$ is approximately a root of the cubic equation \eref{eq:eq14c} with $\delta=0$, we assume $a_{23}=a_{23}^0+\epsilon,~|\epsilon|<<|a_{23}^0|$ where $a_{23}^0=-(k_2/k_1-\alpha_2+2\kappa-\alpha_2\kappa)/(\alpha_2\kappa)$ and use it to convert the cubic equation to the following form:
\begin{eqnarray}
	\alpha_2\alpha_3\left(a_2-a_{23}\right)\left(a_2^2+B_1a_2+B_0\right)+\nonumber \\\frac{\delta}{\kappa}\alpha_2\left(a_2^2-a_{23}a_2-\frac{1}{\alpha_2}\right)\approx 0, \nonumber \\
B_1=\frac{1-\alpha_3}{\alpha_3}+c, ~B_0=\frac{1-\alpha_3}{\alpha_2\alpha_3 a_{23}^0},~\epsilon=\frac{1+\alpha_3}{\alpha_2\alpha_3 a_{23}^0}, \label{eq:eq14f}
\end{eqnarray}
where coefficients $B_1$ and $B_0$ were obtained by expanding \eref{eq:eq14c} after substituting $a_{23}=a_{23}^0+\epsilon$ and neglecting terms of order $\epsilon^2$ and higher. Notice that for a stiff middle resonator $a_{23}^0<0$ as $k_2/k_1-\alpha_2>>0$. Furthermore, one of the roots of the quadratic equation attached with stiffness perturbation is also approximately $a_{23}$ as magnitude of sum of roots is much higher than the magnitude of product of roots ($|a_{23}|>>1/\alpha_2$) for weak coupling. This is expected since the farthest eigenvalue ($\omega_3^2$) is unaffected by perturbation. The other root approximately equals $-1/(\alpha_2 a_{23})\approx -1/(\alpha_2 a_{23}^0)$. Hence the above equation can be converted to the following form:
\begin{eqnarray}
	\alpha_2\alpha_3\left(a_2-a_{23}\right) \nonumber \\
	\resizebox{0.41\textwidth}{!}{$\left(a_2^2+\left(\frac{1-\alpha_3+\delta/\kappa}{\alpha_3}+\frac{1+\alpha_3}{\alpha_2\alpha_3 a_{23}^0} \right)a_2+\frac{1-\alpha_3+\delta/\kappa}{\alpha_2\alpha_3 a_{23}^0} \right)\approx 0.$} \label{eq:eq14g}
\end{eqnarray}
The roots of the quadratic equation in $a_2$ are given by:
\begin{eqnarray}
	a_{21},~a_{22}\approx \frac{1}{2}\left(-\frac{1-\alpha_3+\delta/\kappa}{\alpha_3}-\frac{1+\alpha_3}{\alpha_2\alpha_3 a_{23}^0}\right)\pm \nonumber \\
	\resizebox{0.41\textwidth}{!}{$\frac{1}{2}\sqrt{\left(\frac{1-\alpha_3+\delta/\kappa}{\alpha_3}+\frac{1+\alpha_3}{\alpha_2\alpha_3 a_{23}^0} \right)^2-4\frac{1-\alpha_3+\delta/\kappa}{\alpha_2\alpha_3 a_{23}^0}}.$}  \label{eq:eq14h}
\end{eqnarray}
As expected from \eref{eq:eq5}, the two roots show veering in proximity of $\delta/\kappa=\alpha_3-1$ and change in them due to stiffness perturbation is nonlinear. Far away from this point (outside veering zone), the root with higher magnitude ($a_{22}$ for $\delta/\kappa>0$) can be approximated as the following after neglecting the terms of order $1/(a_{23}^0)^2$ and higher:
\begin{eqnarray}
	a_{22}\approx-\frac{1-\alpha_3+\delta/\kappa}{\alpha_3}-\frac{1}{\alpha_2\alpha_3 a_{23}^0} \label{eq:eq14i}
\end{eqnarray}
Recognizing that the value of $a_{22}<<a_{23}^0$ for $\delta<1$ and using \eref{eq:eq14a}, $a_{32}$ can be obtained.
\begin{eqnarray}
	a_{32}\approx -\alpha_2 a_{23} a_{22} -1. \label{eq:eq14j}
\end{eqnarray}
Substituting approximate value of $a_{22}$ from \eref{eq:eq14i} into \eref{eq:eq14j}, and $a_{23}\approx \alpha_2 a_{23}^0$, we obtain:
\begin{eqnarray}
	a_{32}\approx \alpha_2 a_{23}^0 \left(\frac{1-\alpha_3+\delta/\kappa}{\alpha_3}+\frac{1}{\alpha_2\alpha_3 a_{23}^0} \right) -1. \label{eq:eq14k}
\end{eqnarray}
As $a_{23}^0$ also varies as $\sim 1/\kappa$ (see \eref{eq:eq14e}), variation in $a_{32}$ due to stiffness perturbation is further amplified ($a_{32}\sim 1/\kappa^2$). Sensitivity of  $2_{nd}$ mode (the mode with fast changing modal amplitude ratio) in the three-DOF system is given by:
\begin{equation}
	S_2=\frac{da_{32}}{d\delta}\approx -\frac{k_2/k_1-\alpha_2+(2-\alpha_2)\kappa}{\alpha_3\kappa^2}. \label{eq:eq14l}
\end{equation}
The above expression matches with \eref{eq12} on neglecting the contribution of coupling in numerator (that is, neglecting $(2-\alpha_2) \kappa$).

\begin{figure}[!h]
	\center
	\includegraphics[width=8cm]{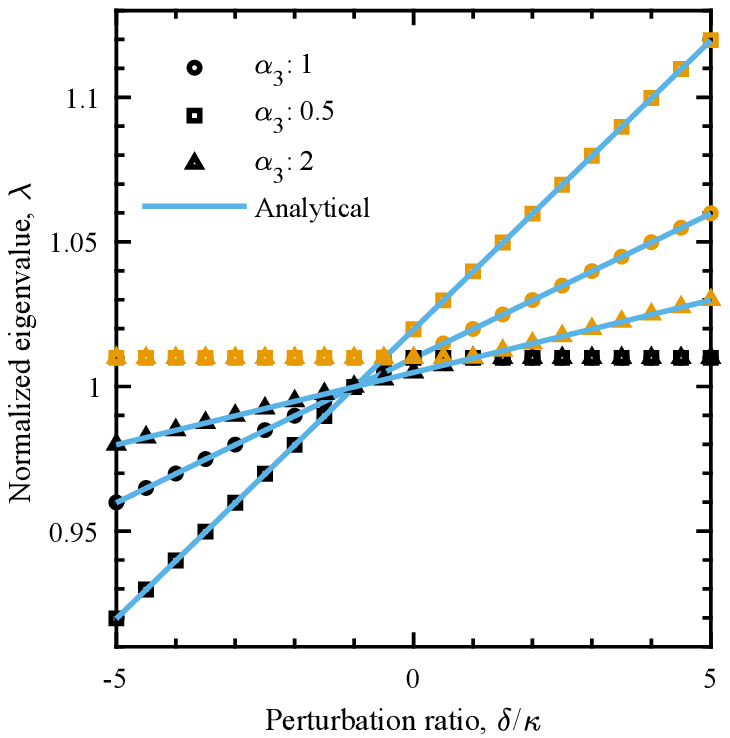}\\
	\includegraphics[width=8cm]{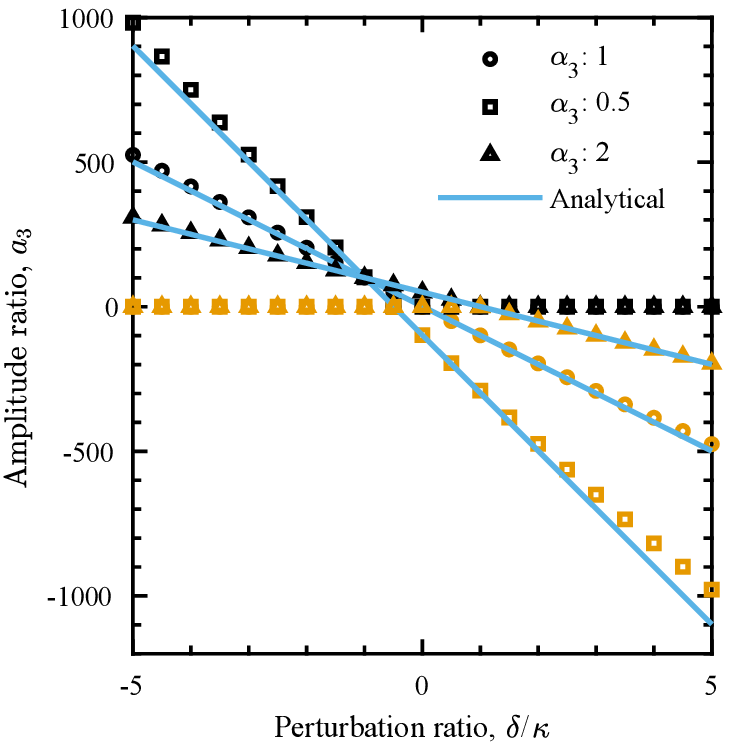}\\
	\caption{Variation of eigenvalues and amplitude ratios with perturbation ratio for the modes undergoing veering for $\alpha_2=1$, and three values of $\alpha_3$ (black and yellow markers correspond to two veering modes, and solid lines are approximate solution from \eref{eq:eq14k}). In the plot, $\kappa=0.01$, and $a_{23}^0=-100$. Notice that with decreasing $\alpha_3$ amplitude ratio undergoes sharper change due to stiffness perturbation.}
	\label{fig:lambda_sensitivity_a3vary}
\end{figure}

\fref{fig:lambda_sensitivity_a3vary} shows the change in modal amplitude ratio $a_3$ with perturbation ratio ($\delta/\kappa$) for modes undergoing veering for $\alpha_2=1$, and three different values of $\alpha_3$.
Note that with decreasing $\alpha_3$ we get a larger change in modal amplitude ratio for the same stiffness perturbation (hence a larger mode sensitivity). Furthermore, for $\alpha_2=1, ~\alpha_3=1$, at zero stiffness perturbation, $a_3$ shows nonlinearity. In literature, an initial bias is suggested to be applied in order to avoid this nonlinear part~\cite{zhao2016mems}. We find that the nonlinear part can be avoided by simply shifting the magnitude of $\alpha_3$ from $1$, eliminating the need for an initial bias. Furthermore, the curves for one particular mode corresponding to all $\alpha_3$ values in the normalized eigenvalue plot as well as in amplitude ratio plot in \fref{fig:lambda_sensitivity_a3vary} pass through a common point. By rearranging the expression for $a_{22}$ in \eref{eq:eq14i}, it is found that this occurs at the value of $\delta/\kappa$ at which the dependence of $a_{22}$ on $\alpha_3$ is nullified, i.e. $\delta/\kappa\approx-1-1/(\alpha_2 a_{23}^0)$.

\begin{figure}[!h]
	\center
	\includegraphics[width=8cm]{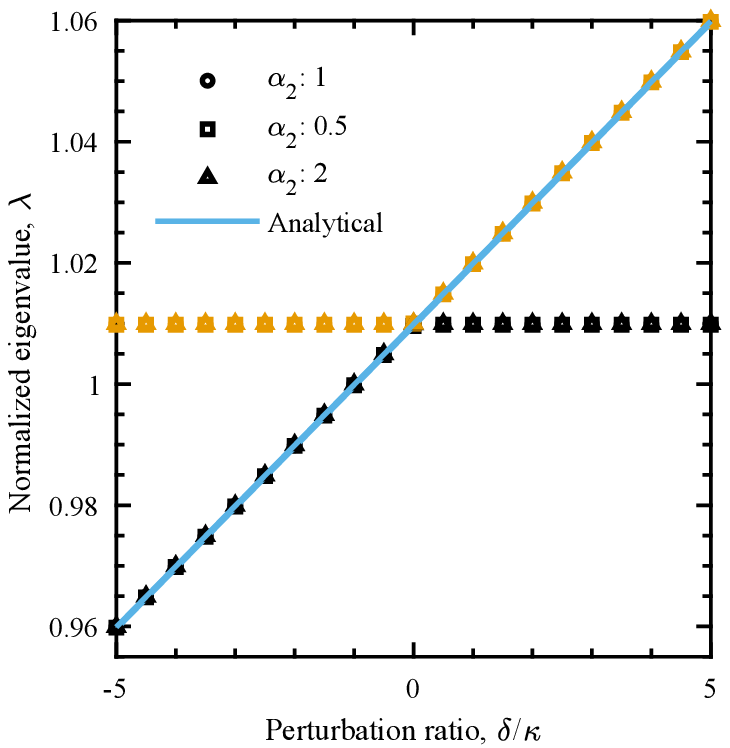}\\
	\includegraphics[width=8cm]{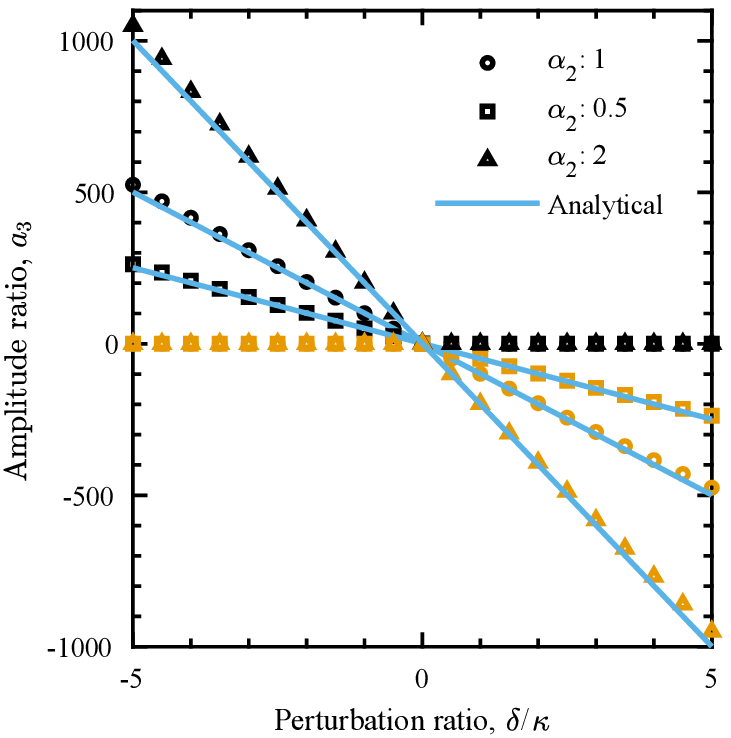}\\
	\caption{Variation of normalized eigenvalues and amplitude ratio with perturbation ratio for the modes undergoing veering for $\alpha_3=1$, and three values of $\alpha_2$ ((black and yellow markers correspond to two veering modes, and solid lines are approximate solution from \eref{eq:eq14k})). For the plot, $\kappa=0.01$, and $a_{23}^0=-100$. Observe that changing $\alpha_2$ does not affect veering eigenvalues. Increasing $\alpha_2$ leads to sharper change in amplitude ratio due to stiffness perturbation.}
	\label{fig:lambda_sensitivity_a2vary}
\end{figure}

\begin{figure}[!h]
	\center
	\includegraphics[width=8cm,angle=90,origin=c]{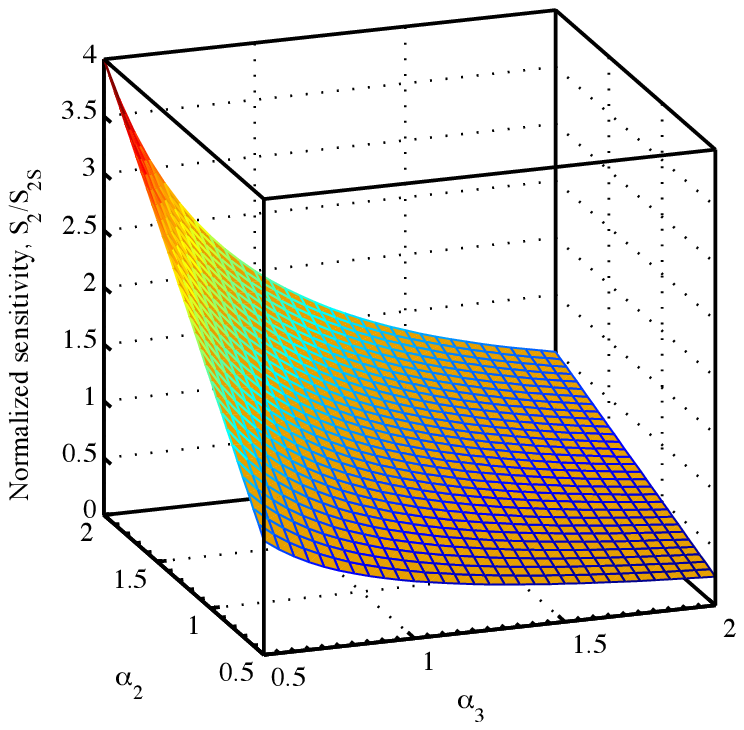}
	\caption{Variation of the ratio of sensitivity with the sensitivity in the symmetric case as asymmetry parameters are varied. For the plot, $\kappa=0.01$, and $a_{23}^0=-100$.}
	\label{fig:normalizedSensitivity_surface}
\end{figure}

\fref{fig:lambda_sensitivity_a2vary} shows change in $a_3$ with $\delta/\kappa$ for modes undergoing veering for $\alpha_3=1$, and three values of $\alpha_2$.
We notice that the eigenvalues do not change, however the modal amplitude ratio decreases with decreasing $\alpha_2$. It can be observed that $a_{32}$ changes faster with stiffness perturbation if we decrease $\alpha_3$ or increase $\alpha_2$, if $k_2$ is varied such that $a_{23}^0$ remains unchanged due to change in $\alpha_2$. Keeping $a_{23}^0~(\approx (\lambda_3-1)/\kappa$) constant ensures that distance between veering eigenvalues and the other eigenvalue remains the same, allowing us to separate the effect of distance between eigenvalues and the effect of change in mass. We conclude that having stiffer and bulkier middle resonator while keeping the natural frequency of the resonator unchanged, also leads to improved sensitivity. In \fref{fig:normalizedSensitivity_surface}, effect of asymmetry on sensitivity is shown. Notice that $\alpha_3$ has a stronger effect on sensitivity.

In the next section, we derive limits of perturbation within which the above sensitivity expression is valid.

\subsection{Perturbation range}
The variation of the amplitude ratio $a_3$ with perturbation deviates from linearity for very small perturbation as well as for large perturbation. So, we find two limits of stiffness perturbation within which amplitude ratio variation with the stiffness perturbation is linear with the maximum deviation from linearity equal to $\gamma$. We focus on the case of $\delta/\kappa>0$.

\subsubsection{Lower limit}
To find the lower limit, we reexamine the value of $a_{22}$ approximated from \eref{eq:eq14h}. For simplicity, we measure perturbation from the proximity of veering, that is $\delta/\kappa=\alpha_3-1$:
\begin{equation}
\delta=\delta_0+\bar{\delta},~\delta_0=\kappa (\alpha_3-1),~\delta_0=0~{\rm for}~\alpha_3=1.
\label{eq:eq14n}
\end{equation}
On changing variable, \eref{eq:eq14h} transforms to give:
\begin{eqnarray}
a_{22}\approx \frac{1}{2}\left(-\frac{\bar{\delta}}{\alpha_3 \kappa}-\frac{1+\alpha_3}{\alpha_2\alpha_3 a_{23}^0}\right)- \nonumber \\
\frac{1}{2}\sqrt{\left(\frac{\bar{\delta}}{\alpha_3 \kappa}+\frac{1+\alpha_3}{\alpha_2\alpha_3 a_{23}^0} \right)^2-4\frac{\bar{\delta}}{\alpha_2\alpha_3 \kappa a_{23}^0}}.  \label{eq:eq14p}
\end{eqnarray}
In comparison to \eref{eq:eq14i}, by taking one more term in the approximation, $a_{22}$ can be written as:
\begin{eqnarray}
	a_{22}\approx-\frac{\bar{\delta}}{\alpha_3\kappa}-\frac{1}{\alpha_2\alpha_3 a_{23}^0}-\frac{\kappa}{\alpha_2^2(a_{23}^0)^2\bar{\delta}}. \label{eq:eq14q}
\end{eqnarray}
Using \eref{eq:eq14a}, $a_{32}$ is approximated to be:
\begin{align}
a_{32}&=(-\alpha_2 a_{23}^0+\alpha_2 a_{22}) a_{22}-1 \nonumber \\
&\approx \frac{\alpha_2 a_{23}^0 \bar{\delta}}{\alpha_3 \kappa}+\frac{1-\alpha_3}{\alpha_3}+\frac{\kappa}{\alpha_2 a_{23}^0 \bar{\delta}}+\frac{2\bar{\delta}}{\alpha_3^2 a_{23}^0\kappa} \nonumber \\
&\approx\frac{\alpha_2 a_{23}^0 \bar{\delta}}{\alpha_3 \kappa}+\frac{1-\alpha_3}{\alpha_3}+\frac{\kappa}{\alpha_2 a_{23}^0 \bar{\delta}},
\label{eq:eq14r}
\end{align}
on neglecting terms of order $1/(a_{23}^0)^2$ and higher, and of oder $\bar{\delta}^2$. Note that nonlinearity originates from the last term with $\bar{\delta}$ in denominator. Deviation of amplitude ratio from linearity should be $\le\gamma$.
\begin{equation}
\left|\frac{\kappa}{\alpha_2 a_{23}^0 \bar{\delta}}\right|\left|\frac{\alpha_2 a_{23}^0 \bar{\delta}}{\alpha_3 \kappa}+\frac{1-\alpha_3}{\alpha_3}\right |^{-1}\le\gamma.
\label{eq:eq14s}
\end{equation}
On simplifying, it gives an inequality which is quadratic in $\bar{\delta}/\kappa$:
\begin{equation}
\left|\left(\frac{\bar{\delta}}{\kappa}\right)^2+\frac{(1-\alpha_3)}{\alpha_2 a_{23}^0}\left(\frac{\bar{\delta}}{\kappa}\right)\right|-\frac{\alpha_3}{\alpha_2^2 (a_{23}^0)^2 \gamma}\ge 0. \label{eq:eq14t}
\end{equation}
The first term, quadratic in $\bar{\delta}/\kappa$, is the dominant term. Solving the resultant quadratic inequality gives:
\begin{equation}
\frac{\bar{\delta}}{\kappa}\ge\frac{\alpha_3-1}{2\alpha_2 a_{23}^0}+ \sqrt{\left(\frac{\alpha_3-1}{2\alpha_2 a_{23}^0}\right)^2+\frac{\alpha_3}{\alpha_2^2 (a_{23}^0)^2\gamma}}. \label{eq:eq14u}
\end{equation}
In terms of actual perturbation, the lower limit is:
\begin{align}
\frac{\delta}{\kappa}\ge & (\alpha_3-1)+\frac{\alpha_3-1}{2\alpha_2 a_{23}^0}+ \sqrt{\left(\frac{\alpha_3-1}{2\alpha_2 a_{23}^0}\right)^2+\frac{\alpha_3}{\alpha_2^2 (a_{23}^0)^2\gamma}}, \nonumber \\
\gtrsim & \alpha_3-1+\frac{1}{\alpha_2|a_{23}^0|}\sqrt{\frac{\alpha_3}{\gamma}}.
 \label{eq:eq14v}
\end{align}
For $\alpha_3=1$, this simplifies to:
\begin{equation}
\frac{\delta}{\kappa}\ge \frac{1}{\alpha_2|a_{23}^0|\sqrt{\gamma}}. \label{eq:eq14w}
\end{equation}
For $\alpha_3=\alpha_2=1$, this expression matches with the expression for nonlinearity of amplitude ratio without damping in an earlier work\cite{zhao2016mems}.

\subsubsection{Upper limit}
To find the upper limit, we make use of the approximation based on block coupled resonators as described in Sec.~\ref{N-DOF_analysis}. Using \eref{eq:eq5} and \eref{eq7}, $a_{22}$ can be obtained:
\begin{equation}
	a_{22}=1-\left(\frac{\lambda_2}{\omega_0^2}-1\right)/\kappa\approx \frac{\alpha_3-1}{\alpha_3}-\frac{\delta}{\alpha_3\kappa}. \label{eq:eq14w}
\end{equation}
We use \eref{eq9} to obtain the expression for $a_{32}$:
\begin{align}
	a_{32}=&(-\alpha_2 a_{23}^0+\alpha_2 a_{22}) a_{22}-1 \nonumber \\
	\approx &-\alpha_2 a_{23}^0\left(\frac{\alpha_3-1}{\alpha_3} \right)+\alpha_2\left(\frac{\alpha_3-1}{\alpha_3} \right)^2-1+ \nonumber \\
	&\left(\frac{\alpha_2 a_{23}^0}{\alpha_3 \kappa}-2\alpha_2\frac{\alpha_3-1}{\alpha_3^2\kappa}\right) \delta+\frac{\alpha_2}{\alpha_3^2\kappa^2}\delta^2 \nonumber \\
	\approx &\underbrace{\left[-\alpha_2 a_{23}^0\left(\frac{\alpha_3-1}{\alpha_3} \right)+\frac{\alpha_2 a_{23}^0}{\alpha_3 \kappa}\delta\right]}_{\approx~{\rm linear~term~in~\eref{eq:eq14r}}}+\nonumber \\
	&\left[\alpha_2\left(\frac{\alpha_3-1}{\alpha_3} \right)^2-1-2\alpha_2\frac{\alpha_3-1}{\alpha_3^2\kappa} \delta+\frac{\alpha_2}{\alpha_3^2\kappa^2}\delta^2\right], \label{eq:eq14x}
\end{align}
where linear term is taken the same as in the calculation of the lower limit in \eref{eq:eq14r}. Enforcing deviation of $a_{32}$ from the first two terms in \eref{eq:eq14x} to be $\le \gamma$ gives:
\begin{eqnarray}
\left|\alpha_2\left(\frac{\alpha_3-1}{\alpha_3} \right)^2-1-2\alpha_2\frac{\alpha_3-1}{\alpha_3^2\kappa} \delta+\frac{\alpha_2}{\alpha_3^2\kappa^2}\delta^2\right|\nonumber \\
\left|-\alpha_2 a_{23}^0\left(\frac{\alpha_3-1}{\alpha_3} \right)+\frac{\alpha_2 a_{23}^0}{\alpha_3 \kappa} \delta \right|^{-1}\le \gamma.
\end{eqnarray}
We solve the above for values of $\alpha_3$ close to $1$, in which case modulus can be removed to obtain:
\begin{eqnarray}
\resizebox{0.41\textwidth}{!}{$\left(\alpha_2\left(\frac{\alpha_3-1}{\alpha_3} \right)^2-1-2\alpha_2\frac{\alpha_3-1}{\alpha_3^2} \frac{\delta}{\kappa}+\frac{\alpha_2}{\alpha_3^2}\left(\frac{\delta}{\kappa}\right)^2\right)$} \nonumber \\
\frac{\alpha_3}{\alpha_2|a_{23}^0|}\left(-(\alpha_3-1)+\frac{\delta}{\kappa} \right)^{-1}\le \gamma,
\end{eqnarray}
which on simplifying yields the following inequality:
\begin{eqnarray}
\frac{1}{\alpha_3 |a_{23}^0|\gamma}\left(\frac{\delta}{\kappa}\right)^2- \left(1+2\frac{\alpha_3-1}{\alpha_3 |a_{23}^0|\gamma}\right)\frac{\delta}{\kappa}+\nonumber \\
\left(\left(\alpha_3-1\right)+\frac{(\alpha_3-1)^2}{\alpha_3 |a_{23}^0| \gamma}-\frac{\alpha_3}{\alpha_2 |a_{23}^0|\gamma}\right)\le 0.
\end{eqnarray}
Solving the above gives the upper limit of perturbation:
\begin{equation}
\frac{\delta}{\kappa}\le \frac{\alpha_3 |a_{23}^0|\gamma}{2}\left(1+2\frac{\alpha_3-1}{\alpha_3 |a_{23}^0|\gamma}+\sqrt{1+\frac{4}{\alpha_2 (a_{23}^0)^2\gamma^2}}\right).
\end{equation}
For $\alpha_3=1$, this simplifies to:
\begin{equation}
\frac{\delta}{\kappa}\le \frac{|a_{23}^0d|\gamma}{2}\left(1+\sqrt{1+\frac{4}{\alpha_2 (a_{23}^0)^2\gamma^2}}\right).
\end{equation}

So, the range of perturbation for which amplitude ratio $a_{32}$ is linear (nonlinearity $\le \gamma$) is given by:
\begin{eqnarray}
\alpha_3-1+\frac{1}{\alpha_2|a_{23}^0|}\sqrt{\frac{\alpha_3}{\gamma}} \lesssim \frac{\delta}{\kappa} \le \nonumber \\
\frac{\alpha_3 |a_{23}^0|\gamma}{2}\left(1+2\frac{\alpha_3-1}{\alpha_3 |a_{23}^0|\gamma}+\sqrt{1+\frac{4}{\alpha_2 (a_{23}^0)^2\gamma^2}}\right).
\end{eqnarray}
Total perturbation range $\Delta_r$ is:
\begin{align}
\Delta_r &=\left(\frac{\delta}{\kappa}\right)_{max}-\left(\frac{\delta}{\kappa}\right)_{min} \nonumber \\ &=\frac{\alpha_3 |a_{23}^0|\gamma}{2}\left(1+\sqrt{1+\frac{4}{\alpha_2 (a_{23}^0)^2\gamma^2}}\right)-\frac{1}{\alpha_2|a_{23}^0|}\sqrt{\frac{\alpha_3}{\gamma}}. \label{eq:range}
\end{align}

\begin{figure}[!h]
	\center
	\includegraphics[width=8cm]{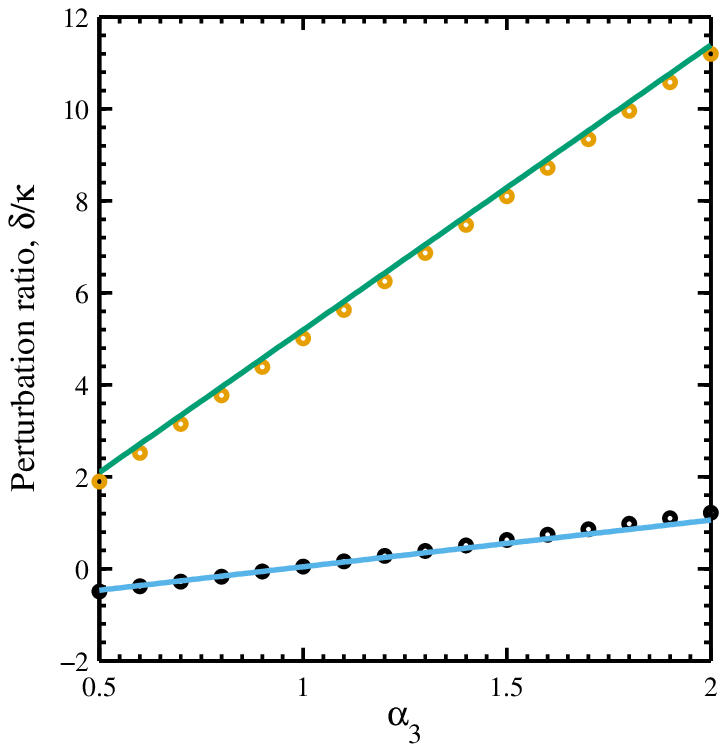}\\
	\includegraphics[width=8cm]{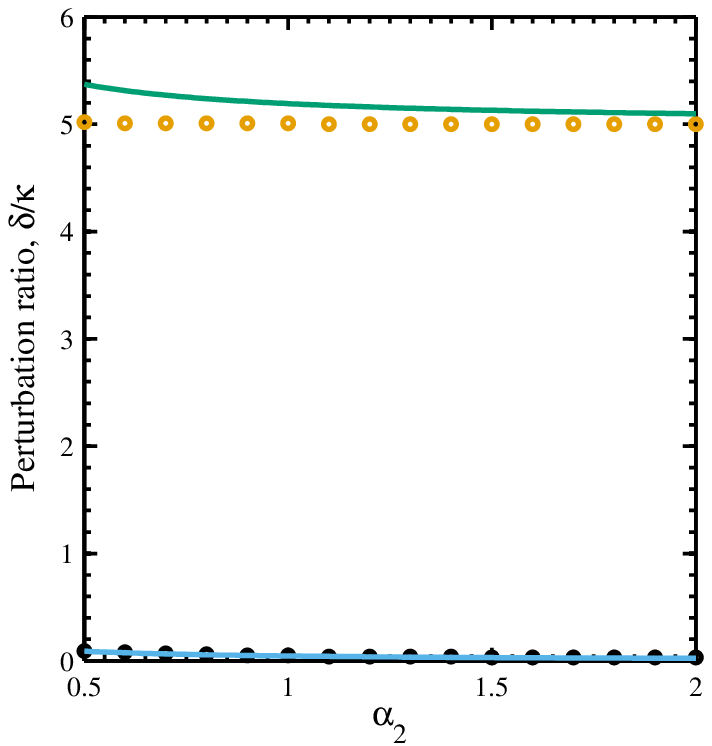}\\
	\caption{Lower and upper limits of perturbation within which amplitude ratio ($a_{3}/a_1$) for the veering mode is linear. Analytical results are shown by continuous lines and circles denote numerical result. The upper plot shows the variation of the perturbation limits with $\alpha_3$ (assuming $\alpha_2=1$) and the lower with $\alpha_2$ (assuming $\alpha_3=1$). For the plot, $\kappa=0.01$, $a_{23}^0=-100$, and acceptable deviation from linearity $\gamma=0.05$.}
	\label{fig:perturbationRange_a3a2vary}
\end{figure}

\begin{figure}[!h]
	\center
	\includegraphics[width=8cm]{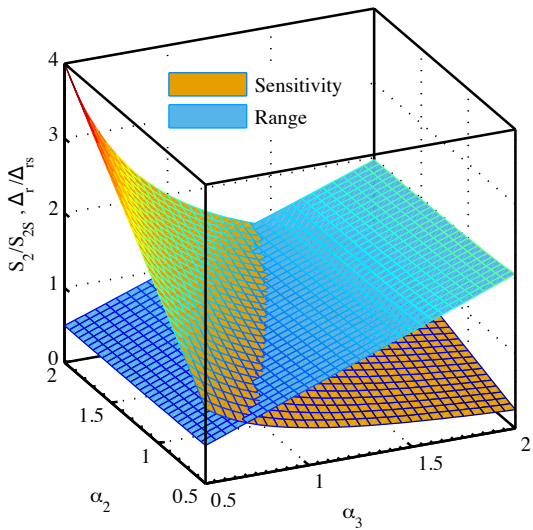}
	\caption{Variation of the ratio of sensitivity and total perturbation range with the sensitivity and total perturbation range in the symmetric case, respectively, as asymmetry parameters are varied. For the plot, $\kappa=0.01$, $a_{23}^0=-100$, and  $\gamma=0.05$.}
	\label{fig:normalizedSpR_surface}
\end{figure}

\begin{figure}[!h]
	\center
	\includegraphics[width=8cm]{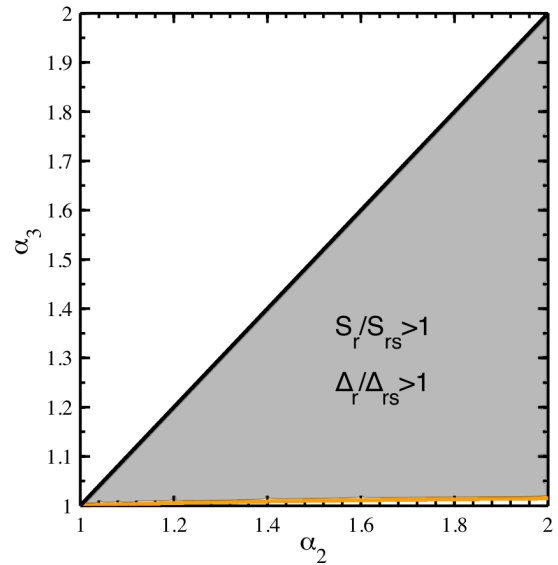}
	\caption{Range of $\alpha_2$ and $\alpha_3$ values for which sensitivity and total perturbation range both is higher than the symmetric system. For the plot, $\kappa=0.01$, $a_{23}^0=-100$, and  $\gamma=0.05$. Note that the origin is the point of symmetry ($\alpha_3=\alpha_2=1$).}
	\label{fig:al3_al2_high_SpRe}
\end{figure}

\fref{fig:perturbationRange_a3a2vary} compares the analytically obtained perturbation range with numerical values. There is a good match between analytical and numerical values. \fref{fig:normalizedSpR_surface} compares the effect of asymmetry on sensitivity and total perturbation range. The plots suggest that the effect of $\alpha_2$ variation on perturbation limits is minimal, even though increasing $\alpha_2$ improves sensitivity. Also, decreasing $\alpha_3$ decreases perturbation range but it improves sensitivity. So, there is a trade-off between sensitivity and perturbation range. However, in a range of values of $\alpha_2$ and $\alpha_3$, sensitivity and the total perturbation range both increase (\fref{fig:al3_al2_high_SpRe}). Asymmetry improves sensitivity and range within the shaded region of this figure. The useful range, in practice, is governed by the smallest amplitude of vibration that can be detected without entering the nonlinear vibration regime.

\section{Conclusion}
\label{conclusion}
Mode localization in a system of weakly coupled resonators has been analyzed using an energy based analytical  approach and recursive relations for modal amplitudes under synchronous motion. The modal recursive relations have been used to derive approximate expressions for sensitivity in~\eref{eq12} and \eref{eq:eq14l}, and for linear range of response in~\eref{eq:range}. These analytical expression are found to  agree with existing results in the literature, where available. Our analysis of localization reveals that carefully engineered asymmetry can enhance the sensitivity and linear range of response for sensing applications that rely on mode localization. Practical demonstration of these results in MEMS sensors remains for future work.

\begin{acknowledgments}
We gratefully acknowledge the support from Natural Sciences and Engineering Research Council of Canada (NSERC).
\end{acknowledgments}

\bibliography{references}

\end{document}